# DNA Scaffold Carrier Assisted Protein Translocation Through Solid-State Nanopore

Jing Yang[1,2*], Juan Wang[2‡], Ranfeng Wu[3], Yiming Chen[2], Cheng Zhang[2*]

[1]School of Control and Computer Engineering, North China Electric Power University, Beijing 102206, China

[2]School of Electronics Engineering and Computer Science, Peking University, Beijing 100871, China

[3]School of computer science and technology, Dalian university of technology, 116024, China

[‡] The authors contributed equally to the first author

*Corresponding author Email: zhangcheng369@pku.edu.cn; yjzcdd_2000@ncepu.edu.cn

**Abstract:**

The detection of biomolecules at the single molecular level have important applications in the fields of biosensing and biomedical diagnosis. Solid state nanopore (SS-nanopore) is an effective tool to perform the single molecular detection, due to its unique properties of label-free and less sample consumption. The current SS-nanopore translocations of small biomolecules are driven by the electronic field force, thus easily influenced by the charges, shapes and sizes of the target molecules. Therefore, it remains a great challenge to control biomolecules to translocate through SS-nanopore, particularly for the protein with complex conformations and different charges. Towards this problem, we developed a DNA scaffold carrier assisted strategy to help protein translocation through SS-nanopore, thus facilitating the target protein detection. Taking advantages of a DNA carrier loading effect, the current signal to noise ratios are significantly improved. Our proposed method is sensitive, convenient and can detect real-time sample nanopore translocations. This method opens up a wide range of applications for SS-nanopore detection and promote the development of single molecular detections.

**Introduction**

Solid-state nanopore (SS-nanopore) is a recently developed single molecular detection method with the properties of high throughput and sensitivity[1-4]. Taking advantages of the label-free, sensitive and real-time features, SS-nanopore has been widely applied to perform the single molecular detection. Recently, a wide range of biomolecules have been studied with SS-nanopores by the specific translocations, including proteins[5-10], drugs[11-16], nucleotides[17-21] and polysaccharides[22]. In fact, the current SS-nanopore translocations of small biomolecules are driven by the electronic field force, thus easily influenced by the electronic charges, shapes and sizes of the target molecules. Therefore, it remains a great challenge to control biomolecules to translocate through SS-nanopore, particularly for the protein with complex conformations and different charges.

On the other hand, many DNA assembly methods have been developed to assist the SS-nanopore translocation of biomolecules, for example, one-dimensional linear DNA carrier assisted nanopore detections is established to escort biomolecular targets through the SS-nanopore[23-26]. With the developments of DNA self-assembly, a variety of complex two- or three-dimensional DNA structures have been designed to assist the SS-nanopore translocation[27-36], mainly focusing on the structural factors induced the signal changes of the nanopore translocation. In particular, polyhedral DNA structures have been established with unique three-dimensional conformational characteristics and dynamic manipulations. Actually, polyhedral DNA structures may potentially provide an advanced DNA carrier with more complex conformation and manipulations to obtain better signal to noise ratios. Compared to the linear DNA carrier, however, it is rarely reported that using polyhedral DNA carrier to assist biomolecules to translocate SS-nanopore[36-40].

In this study, we reported a DNA scaffold carrier assisted strategy to help protein translocation through SS-nanopore, thus facilitating the target protein detection. Two kinds of polyhedral DNA carriers are designed by using tetrahedral and cube DNA structures, with a side length of about 10.2 nm and 6.8 nm, respectively. By coating protein with polyhedral DNA carrier, the structures and electron charges of the protein are greatly changed, thus significantly affect the carrier assisted protein nanopore translocation process and the corresponding signal. This DNA carrier assisted approach may provide easy and sensitive method for the target protein nanopore detection, and have more applications in biosensing and molecular diagnosis.

**Results**

**DNA carrier design and nanopore sensing**

Fig1a illustrates the schematic diagram of two DNA carrier. There are three kinds of structures (opTET, TET and cubeout) presented in this paper. opTET structures are composed of four DNA single strand, which are s1-in (126nt), S2-in (112nt), S3-Lo-2 (69nt) and S4 (93nt). Based on the opTET structure, a single strand S3-ch (24) is added to the opTET to form a closed tetrahedron. Notably, opTET and TET both have unbound single strands inside the tetrahedron (figura1a, the red part). This does not affect the tetrahedron as a carrier structure. Instead, the subsequent experiments will revolve around introducing modified binding proteins to the single strand. In comparison to TET, opTET has a gap, due to the absence of strand S3-ch, which is complementary with strand S1-in. The gel analysis was utilized to detect two structures TET and opTET. As shown in Figure 1b, the band of TET in line 5 ran slightly slower than that of opTET in line 4. This phenomenon results from the absence of strand S3-ch. Additionally, the AFM images of opTET structure in figure 1d further confirm the formation of opTET.These two samples were recycled through gel recovery for the nanopore detection experiment. Figure 1e shows a feature translation signal diagram of TET at 500mV in a 30nm nanopore. It is clear from the TET characteristic signal is successfully detected through nanopore.

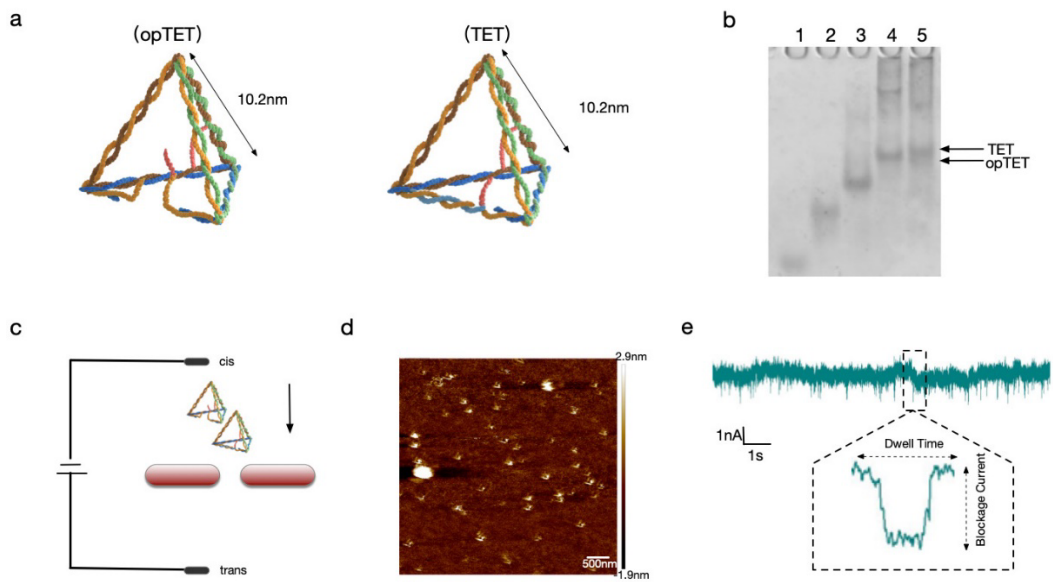

**Figure 1.**(a):an illustration of the open tetrahedron (opTET) and the closed tetrahedron (TET). (b): an illustration of the formation of tetrahedrons. Line 1: S1-in; Line 2: S1-in + S2-in;Line3: S1-in + S2-in + S3-LO-2 ;Line4: open tetrahedron(opTET);Line5: closed tetrahedron(TET).(c):Solid nanopore passing through nanopore in a schematic diagram.(d):opTET structure of AFM.(e): Current track diagram of the TET structure at 500mV, A dotted box represents an enlarged view of the above signal, with dwell time in the horizontal direction and blockage current in the vertical direction.

**Self-assembled tetrahedral through nanopores**

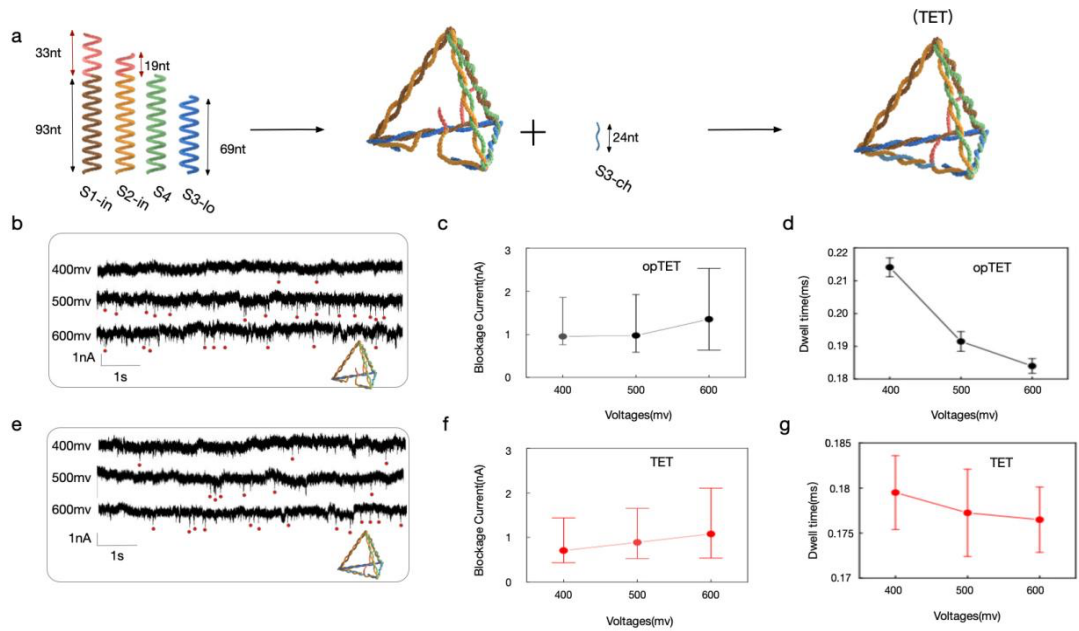

**Figure 2.**(a):schematic diagram of the formation process of open tetrahedron (opTET) and tetrahedron (TET).(b):Electrical signal diagram of opTET at 400mV, 500mV and 600mV, The red dot indicates an event that has taken place in the nanopore.(c):400mV, 500mV and 600mV average current of opTET in 30nm nanopore. d: Average plot of 400mV, 500mV and 600mV passing time of opTET at 30nm nanopore (e):electrical signals of the TET at 400mV, 500mV and 600mV, The red dot indicates an event that has taken place in the nanopore (f):the average current of the TET at 400mV, 500mV and 600mV at 30nm.(g):Average plot of 400mV, 500mV and 600mV passing time of TET at 30nm nanopore.

Next, solid nanopores were used to implement experiments on TET and oPTET structures respectively to detect these structures further. It is worth noting that TET is formed by adding the strand S3-ch, complementary with S1-in, thus there is not much difference between the two structures. For opTET and TET, we have conducted translation experiments under different voltages. According to the experimental results, opTET has a smaller blockage current at 400mV compared to 500mV and 600mV, and there are fewer signals at 400mV than those at 500mV and 600mV (Fig.). As the voltage increased from 400 mV to 600 mV, both the event occurrence frequency and the magnitude of the current blockage increased, while the dwell time decreased .As we

expected，This is because most of the tetrahedron appeared blocking phenomena at the low voltage makes the tetrahedron difficult to pass through. Although the molecules pass through the nanopore under the action of voltage, the dwell time will be longer than that at high voltages. However, when voltage is increased, molecular structure translation becomes easier, therefore opTET and TET display the same trend in blockage current and dwell time with increased voltage. Under different voltages, we also performed gaussian fitting for the translation time of these two structures. The experimental results were verified by fitting data in FIG. S8, the supplementary material. The fitting results of opTET translation time at 400mv, 500mv and 600 mV were 0.21±0.11ms, 0.19±0.11ms and 0.18±0.01ms, respectively. The fitting results of TET were 0.17±0.1ms, 0.17±0.09ms and 0.16±0.08ms, respectively. It is demonstrated that both of the molecular structures can detect translocation signals using solid nanopores, and the number of signals is relatively high.

**Self-assembled tetrahedral carrier structure carries protein through the nanopore**

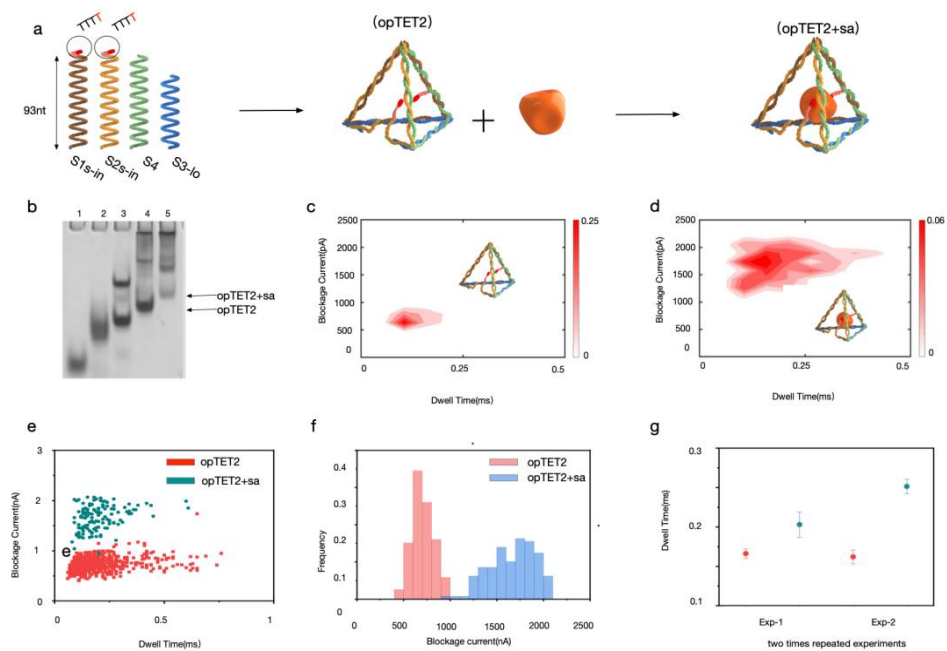

**Figure 3.**(a):schematic diagram of the formation of opTET2 and opTET+sa. The red T represents biotin modification on the base. (b): Schematic diagram of running glue of

opTET2 and opTET2+sa. Line: S1S-in; Line2: S1S-in + S2S-in; Line3: S1s-in+S2s-in+S4.;Line4:opTET2;Line5: opTET2+ sa. (c): Scatter plots of the translocation signals corresponding to the opTET2. (d) Scatter plots of the translocation signals corresponding to the opTET2+sa. (e): Scatter diagram of opTET2 and opTET2+ sa. Red represents opTET2 and blue represents opTET2+ sa. (f):Histograms of the current blockades show the translocation events of opTET and opTET+ sa. Where, red represents opTET2, blue represents opTET2+sa. (g): Two times repeated experiment results of Exp-1, -2, red and blue refer to opTET2 and opTET2+sa respectively.

Because opTET and TET have similar structural features, we speculate if the modification of carrier structure and the binding of specific proteins can distinguish different structures through structural translation events. Based on this, we designed a new carrier structure, called opTET2. Compared with opTET ,opTET2 changed two strands s1-in and S2-in. The figure3a shows two strands S1s-in and S2s-in that have been changed in opTET2. Adding biotin to S1s-in and S2s-in enables streptavidin (sa) to bind to it when added. Through gel result, we determined whether the protein could be successfully embedded into the tetrahedron. As shown in Fig3b, the line 4 was opTET2, and the line 5 was incubated with 5um sa and opTET2 for 2 hours at room temperature. The resulting new structure is called opTET2+sa.According to the results of gel, we can see that opTET2 ran slower after adding sa, which shows that sa was successfully embedded.Next, solid nanopore was used to verify whether the change of structure would lead to the difference in blockage current and dwell time between the two structures. We let opTET+sa and opTET2 pass through solid nanopore respectively, and the experimental results are shown in Fig3c,d.At the beginning, the blockage current of opTET2 was about 600-700Pa, as shown in the results. After sa was introduced, opTET2+sa blockage current increased significantly. This is because compared with the empty opTET2 structure, opTEE2+ sa structure had a larger internal structure due to sa insertion，Passing through the nanopore is difficult.Fig.3e,f shows the dwell time and blockage current of opTET2 and opTET2+sa passing through the nanopore. The sample of opTET2 and opTET2+ sa passing through the nanopore has a volume of 200ul and a concentration of 5nm. Afterwards, the samples were drilled under the same experimental conditions. We listed the characteristic diagrams of the translation signals of the two structures in figs10,12, and the specific translation signals are given in the supplementary material (figs11,13).Through the above experiments, we observed that when opTET2 was modified and sa combined, compared with the opTET2 structure, it could be detected by using solid nanopores and accurately distinguish the two structures.

**Light controlled self-assembled tetrahedral carrier**

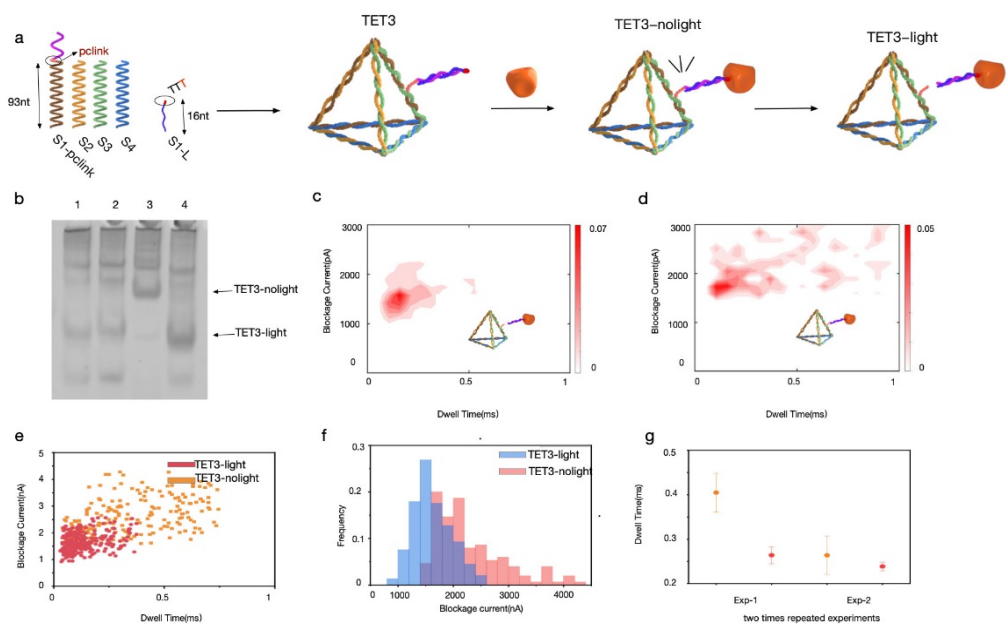

**Figure 4.**(a):Schematic diagram of TET3-nolight and TET3-light formation. The red T represents biotin modification on the base.(b): TET3-nolight and TET3-light run glue diagram, line1: S1-Pclink + S2+S3+S4 ; line2: S1-Pclink +S2+S3+S4+S1-L(TET3); Line3 is to add 5um sa to TET3( TET3-nolight); Line 4: TET3-nolight after ten minutes of ultraviolet irradiation(TET3-light).(c): Scatter plots of the translocation signals corresponding to the TET3-light.(d): Scatter plots of the translocation signals corresponding to the TET3-nolight (e): Through-nanopore time and current scatter diagrams for TET3-light and TET3-noLight, with red indicating the through-nanopore light and yellow indicating the current distribution histogram for TET3-NOLight. (f) : Histograms of the current blockades show the translocation events of TET3-nolight and

TET3-light. The blue represents TET3-nolight, and the red represents TeT3-light.(g): Two times repeated experiment results of Exp-1, -2, orange and red refer to TET3-nolight and TET3-light respectively.

We have validated through the above experiment that opTET2 together with sa will cause the blockage current and dwell time of translation nanopore to rise, but since the biotin and sa combination is strong, it is hard to remove the sa again, so we designed a new structure (TET3) to control the loss of sa on the carrier structure protein.Protein binding will increase blockage current. Experiments have confirmed protein binding increases the blockage current, When the protein is shed, it will cause the blockage current to drop? This new structure is called TET3-nolight, and it incorporates UV light to control the release of proteins. Fig.4a illustrates that the S1-Pclink strand forming the TET3-nolight structure has pclink modification, which is a special modification that is sensitive to UV light(figura4a, the red part on the S1-pclink). Under ultraviolet irradiation, the DNA strand will break from pclink modification base. In the TET3 structure ,the red part of S1-pclink are four bases T and pclink is modified on the fourth base T. Strand S1-L are introduced in the fig4.a, which can hybridize with S1-pclink. In addition,Strand S1-L have the Biotin modification ,which can combine with sa. The new structure combined with sa is called TET3-nolight.When the uv light is introduced, the double-strand structure combined with the sa will break at the position of pclink modification, resulting in TET3-light. TET3-light and TET3-nolight were verified through gel result as shown in fig.5b, but cannot be distinguished in the nanopore. Therefore, The TET3-light and TET3-nolight were detected by nanopores. According to Fig5d, the blockage current and dwell time of TET3-nolight translation the nanopore are relatively high. However, The dwell time and blockage current are relatively concentrated in a range when ultraviolet light is introduced into the structure in fig5c. fig5e shows the scatter diagram analysis of dwell time about the two structures. According to theresult, the data of TET3-light are relatively scattered, and their blockage current and dwell time are higher than those of TET3-nolight.The figure5f shows that the distribution of the blockage current of TET3-nolight is larger. We conducted two independent repeated experiments on the experiment in fig5g. Both experiments showed similar results. The dwell time of the carrier structure of binding protein was longer. In this experiments, we used Ultraviolet irradiation to control protein shedding.Solid nanopore was used to detect whether the carrier structure would reduce the blocking current and dwell time due to protein shedding.The experimental results further verified that the structural changes caused by the combination of carrier structure and protein could be detected through solid nanopore

**Self-assembled cubic carrier structure carries protein through the pore**

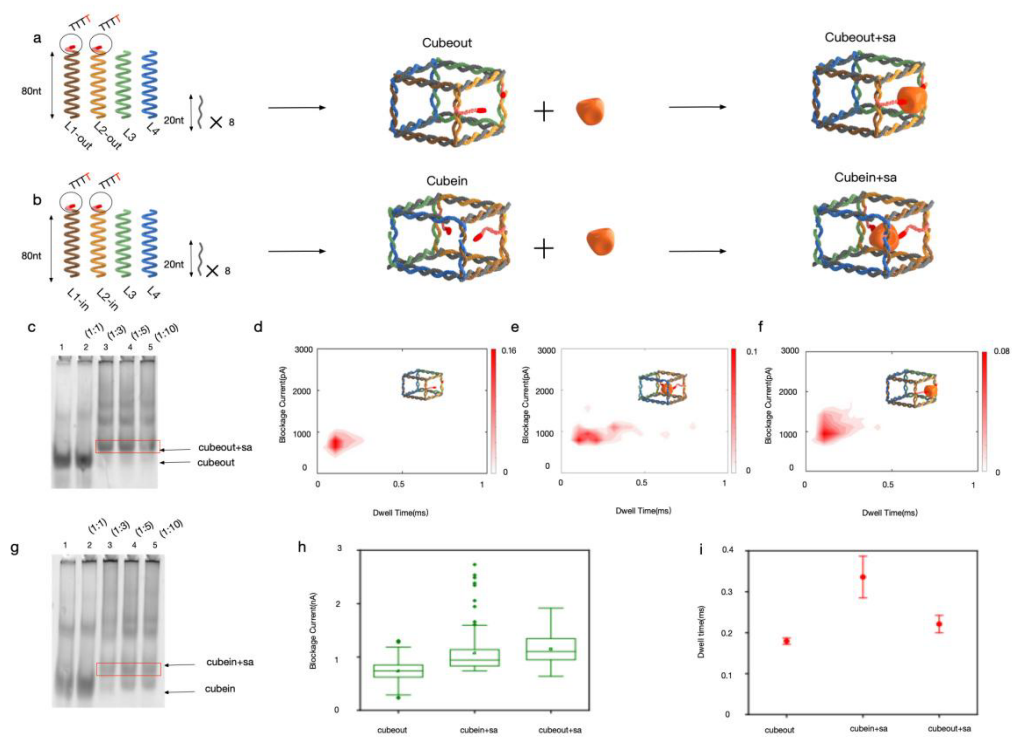

**Figure 5.**(a): schematic diagram of Cubeout and Cubeout +sa structure formation.(b): Schematic diagram of Cubein and Cubein + sa structure formation.(c): cubeout and cubeout+sa gel result, Line1:cubeout; line2: Cubeout added 1um sa;line3: Cubeout added 3um sa; line4: Cubeout added 5um sa;line5: Cubeout added 10um sa. (d): Scatter plots of the translocation signals corresponding to the Cubeout.(e): Scatter plots of the translocation signals corresponding to the cubein + sa.(f): Scatter plots of the translocation signals corresponding to the Cubeout + sa.(g): Cubein and Cubein+sa gel

result, line1: Cubein: line2: Cubein added 1um sa ;line3: Cubein added 3um sa ;lin4: Cubein added 5um sa ; line5: Cubein added 10um sa.(h): Current box diagram of Cubeout and Cubein+ sa and Cubeout+ sa.I: mean through-nanopore time of Cubeout and Cubein+ sa and Cubeout+ SA.

Based on the above experiments, the tetrahedral carrier structure shows good It shows good carrying effect. To verify whether the carrier structure binding of proteins will change the blockage current, we designed two carrier structures.The Cubeout structure is made up of 12 strands, L1-out(84nt), L2-out(84nt),L3(80nt),L4(80nt),D1(18nt),D2(18nt),D3(18nt),D4(18nt),D5(18nt),D6(18nt),D7(18nt),D8 (18nt).The Cubein structure is also made up of 12 strands, L1-in(84nt), L2-in(84nt),L3(80nt),L4(80nt),D1(18nt),D2(18nt),D3(18nt),D4(18nt),D5(18nt),D6(18nt), D7(18nt),D8 (18nt). The sides of two structure are about 6.8nm.L1-in and L2-in and L1-out and L2-out all have biotin modification on the 3 'end base T in Fig5a,b red base.Therefore, the modification sites can bind sa . The difference is that Cubeout binds SA outside and Cubein binds SA inside the structure.The formation of the two structures can be verified by gel results. From this gel result it is clealy that sa with 3um generally reached saturation when combined with these two structures, since the height of the band with 5um sa and 10um sa was basically the same as that of fig5c,g. Therefore, we recovered Cubein+sa and Cubeout+sa by combining sa with a concentration of 3um for nanopore detection experiment. The results show that the blockage current of the cube added with sa increases, which is similar to the effect of opTET2 binding sa. According to the experimental results,Cubeout + sa and Cubein + sa have different changes in blockage current(Fig.5h).Cubein+sa has a larger blockage current than Cubeout+sa.Cubeout-sa is different from Cubein-sa because the carrier structure combined sa are outside .In Cubeout+sa, The interactions between the sa and cube will increase structure side length,which would increase the blockage current .Gaussian fitting of blockage current was also performed for the three structures, where Cubeout (0.73± 0.16nA), Cubeout + sa (1.2± 0.32nA), and Cubein + sa (1.0±

0.38nA)(Figs18).On the dwell time, these three structures are also different(Fig.5i). cubein + sa was a little longer on dwell time, likely because the sa is combined inside the cube,which increases the volume of the structure. It makes it difficult for Cubein+sa pass through nanopoe, resulting in a long dwell time.Through the cube carrier structure binding protein experiment, we also confirmed that the structural changes caused by the carrier structure modifications and binding of proteins can also be detected by solid nanopores. And when the same carrier structure is modified at different positions, the structure that binds to the protein can also be detected.

**Discussion**

In this work, to obtain the better signal to noise ratios when the protein translocated though the SS-nanopore, we developed the DNA tube and tetrahedral carriers with a side length of about 10.2 nm and 6.8 nm respectively, and experimentally demonstrated their advances. At the same time, we also proved that the conformational variations of the polyhedral carriers were consistent to the electronic signal changes. Specifically, the translocation characteristics of the empty carriers were implemented first. As expected, the different carriers generated specific signals and blockage current. Then, to characterize the translocation of the protein included carriers, the biotin modification was carried out on the surfaces of the two structures. Clearly, the results further demonstrated the successful design of our DNA polyhedral carries. On the other hand, the UV-light sensitive modification was introduced to the DNA carriers, and we also verified the including state of the protein could be further controlled by the UV-light. In summary, this work provided a label-free and rapid tool to characterize vector structure-protein binding, and we believe it will motivate a merging of works for the nanopore diagnosis and biosensing.